\documentclass[11pt,preprint,tighten]{aastex}

\newcommand{\zem}{\mbox{$z_{\rm em}$}}

\begin{document}


\title{Discovery of four X-ray quasars behind the Large Magellanic
Cloud\altaffilmark{1}}

\author{A. Dobrzycki\altaffilmark{2}, P. J. Groot\altaffilmark{2,3},
L. M. Macri\altaffilmark{2} and K. Z. Stanek\altaffilmark{2}}

\altaffiltext{1}{Based on observations collected at the Magellan Baade
6.5-m telescope.}

\altaffiltext{2}{Harvard-Smithsonian Center for Astrophysics, 60
Garden Street, Cambridge MA 02138, USA, e-mail:
[adobrzycki,lmacri,kstanek]@cfa.harvard.edu}

\altaffiltext{3}{Present address: Department of Astrophysics,
University of Nijmegen, PO Box 9010, 6500 GL Nijmegen, The
Netherlands, e-mail: pgroot@astro.kun.nl.}

\shorttitle{X-ray quasars behind LMC}
\shortauthors{Dobrzycki et al.}


\begin{abstract}

We present the discovery of four X-ray quasars ($\zem=0.26, 0.53,
0.61, 1.63$) located behind the Large Magellanic Cloud; three of them
are located behind the bar of the LMC. The quasars were identified via
spectroscopy of optical counterparts to X-ray sources found
serendipitously by the Chandra X-ray Observatory satellite. All four
quasars have archival $VI$ photometry from the OGLE-II project; one of
them was found by OGLE to be variable. We present the properties of
the quasars and discuss their possible applications.

\end{abstract}

\keywords{Magellanic Clouds --- quasars: individual 
(OGLE050736.52--684751.7, OGLE050833.29--685427.5,
OGLE050924.05--672124.1, OGLE051853.19--690217.7)
--- X-rays: general}

\slugcomment{Accepted March 6, 2002 for publication in Ap.J.Letters}


\section{Introduction}

There are a number of reasons that make quasars behind the Large
Magellanic Cloud (LMC) --- as well as other nearby galaxies --- very
interesting. Among other, they provide a good inertial reference
system (e.g.\ Anguita, Loyola \& Pedreros 2000) to measure proper
motion of the LMC. Also, such quasars can provide a line-of-sight
probe of the interstellar medium in the LMC (e.g.\ Bowen, Blades \&
Pettini 1995; Haberl et al.\ 2001; Kahabka, de Boer, \& Br\"un
2001). One aspect of such studies is to investigate the dust-to-gas
ratio in galaxies (e.g.\ Fall \& Pei 1989).

There are only a dozen or so of publicly known quasars in the general
direction of the LMC. All these quasars are located away from the bar
of the LMC, in fairly sparse stellar fields. One was found by Blanco
\& Heathcote (1986) via a grism survey. Crampton et al.\ (1997; see
also Kahabka, de Boer, \& Br\"uns 2001) listed several other. In their
analysis of the LMC proper motions, Anguita, Loyola, \& Pedreros
(2000) show a sample of three quasars, including that of Blanco \&
Heathcote (1986).

Drake et al.\ (2001) are currently pursuing a project of the study of
the proper motions using quasars found in MACHO data. They quote a
sample of $\sim$30 quasars in the dense stellar regions of the LMC,
pre-selected using their variability as observed by the MACHO project,
and then confirmed spectroscopically. However, the positions of those
quasars are not publicly available.

A good starting point for searching for quasars is to utilize the fact
that they are often bright in X-rays, and a considerable number of
quasars have been found in that way (including the ones from Crampton
et al.\ 1997). However, the limited spatial resolution of earlier
X-ray missions rendered this method impractical for the dense parts of
the LMC, where a single X-ray source would have many optical
candidates in its error box. The launch of the Chandra X-ray
Observatory (Weisskopf et al.\ 2000) made application of this method
to the LMC possible. Chandra's superb spatial resolution and excellent
positional accuracy significantly reduce the source confusion problem.

Another characteristic that was used in quasar searches was their
irregular variability. Several quasar surveys utilizing variability
were performed or are on-going (e.g.\ Drake et al.\ 2001; Meusinger \&
Brunzendorf 2001; Rengstorf et al.\ 2001).

Between 1997 and 2001, large parts of the Large Magellanic Cloud were
monitored for microlensing events by the Optical Gravitational Lensing
Experiment (OGLE-II: Udalski, Kubiak \& Szyma\'nski 1997). Udalski et
al.\ (2000) released photometry and astrometry of more than seven
million objects from the central parts of the LMC\footnote{Data
available from ftp://bulge.princeton.edu/ogle/ogle2/maps/lmc/}.  In
addition, a large catalog of 68,000 variable objects observed by
OGLE-II in both the LMC and the SMC was prepared by \.Zebru\'n et al.\
(2001)\footnote{Data available from
http://bulge.princeton.edu/$\sim$ogle/ogle2/dia/}, based on a version
of the image subtraction software (Alard \& Lupton 1998) developed by
Wo\'zniak (2000).

The availability of Chandra observations of fields in the LMC,
combined with the availability of the OGLE database, allow combining
the two methods in a search for quasars in the dense regions of the
LMC. The Chandra detectors have a relatively large field of view and
each pointing yields dozens of serendipitous sources.

\section{The archival data}

We searched the Chandra archive for imaging (i.e.\ with no grating)
observations whose pointings would coincide with OGLE fields. At
present, there are three such observations that are publicly
available, ObsIDs 118, 125, and 776. A fourth observation fulfilling
this criterion, ObsID 1991, was kindly provided to us by the P.I.,
K.~Borkowski. All four observations were done with ACIS as the focal
plane detector, in various chip configurations. A single ACIS
observation typically covers 0.12~deg$^2$. There was no overlap
between the observations.

We reduced and analyzed the X-ray data using tools available in the
CIAO 2.2.1 and SHERPA software
packages\footnote{http://cxc.harvard.edu/ciao/}. We cleaned the
electronic streaks in ACIS-S4 chip using {\em destreak}, and we
searched the data for serendipitous point sources using the sliding
cell tool {\em celldetect}; of the three detect tools available in
CIAO this one is the most robust when it comes to the detection of
point sources. We utilized the relation between the off-axis angle and
signal-to-noise threshold from Dobrzycki et al.\ (2000). Overall, we
identified 361 X-ray sources in our four Chandra observations. For
each source, we then identified both the closest OGLE object and the
closest OGLE variable. Some of the data available from the Chandra
archive were processed before the best Chandra pointing calibration
was available, and we used 5~arcsec as the threshold for the position
match. We ended up with a list of 110 OGLE objects meeting those
criteria and they formed a list of candidates for followup
spectroscopy.

\section{Observations}


\begin{figure}[!h]
\plotone{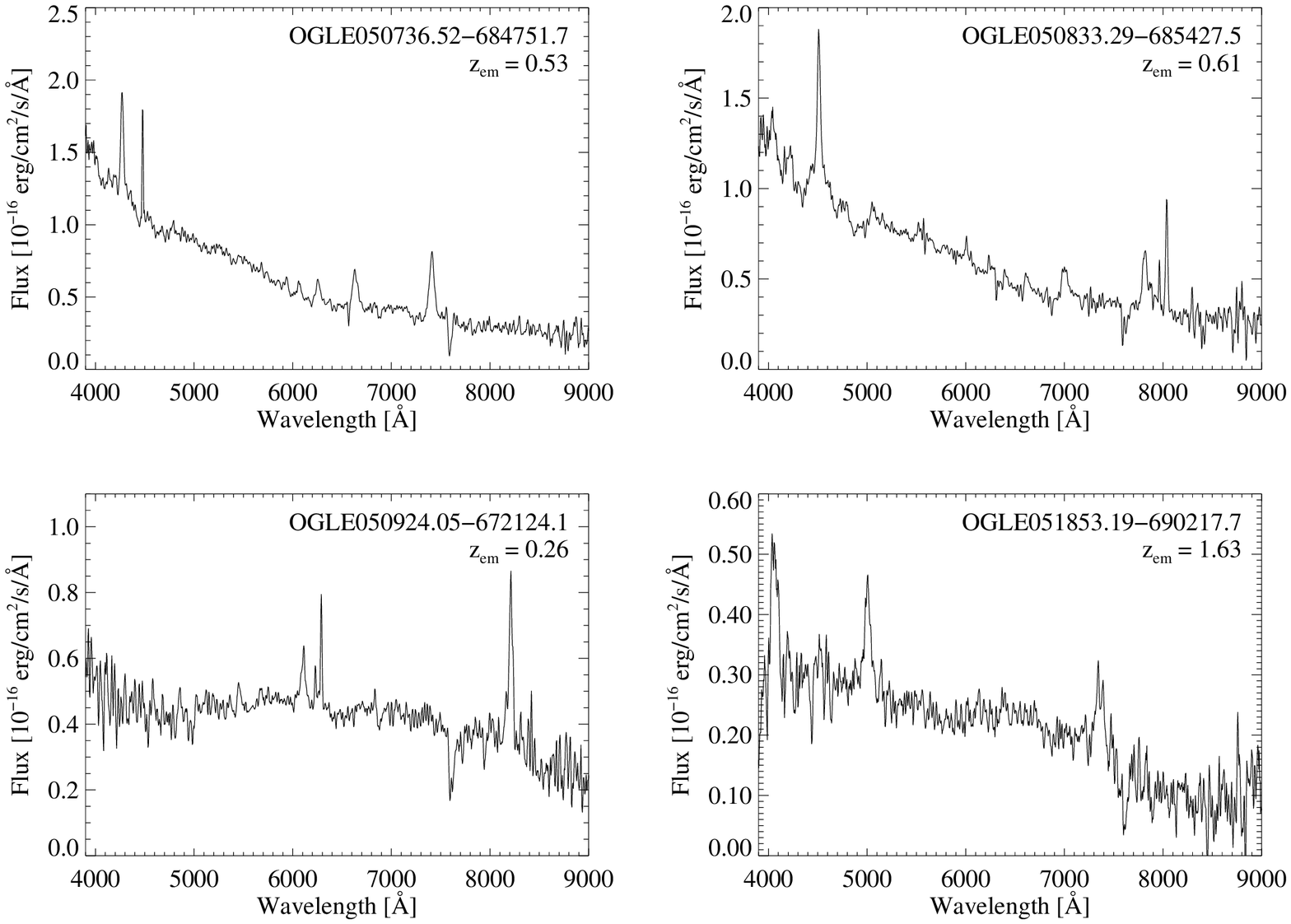}
\caption{Optical spectra of the quasars, taken with the Magellan Baade
6.5-meter telescope. For presentation, the spectra were smoothed with
a 3-pixel box.\label{fig:magellan}}
\end{figure}


The optical spectra were obtained on 2002 January 22-23 with the
Magellan Baade 6-5 meter telescope.  We observed 35 objects out of 110
available candidates; the observed objects were primarily the
brightest in X-rays.

We used the LDSS-2 imaging spectrograph. The instrument uses a
2048$\times$2048 SITe\#1 CCD with a scale of 0.38~arcsec/pixel, a gain
of 1~$e^-/$ADU, and a readout noise of $7e^-$. The slit width was
1.03~arcsec and the grism setting was 300~l/mm, yielding a nominal
resolution of 13.3~\AA. Exposure times ranged from 300 to 600
seconds. All observations were carried out at the parallactic angle.
Additionally, two spectrophotometric standards were observed: LTT 1788
and LTT 4816 (Hamuy et al.\ 1992). Following each observation, a
He--Ne arc lamp spectrum was acquired for wavelength calibration
purposes. Spectra were reduced in the standard way using IRAF.

\section{New quasars}


\begin{deluxetable}{ccccc}
\tabletypesize{\scriptsize}
\tablewidth{0pt}
\tablecaption{Quasar optical data.\label{tab:opt}}
\tablehead{%
\colhead{OGLE ID/Coords.\tablenotemark{a}} &
\colhead{OGLE field} &
\colhead{Redshift} &
\colhead{$V$} &
\colhead{Remarks\tablenotemark{b}} \\
\colhead{} &
\colhead{} &
\colhead{} &
\colhead{[mag]} &
\colhead{}
}
\startdata
050736.52--684751.7 & SC11    & 0.53 & 19.81 &     \\
050833.29--685427.5 & SC11    & 0.61 & 19.00 & (1) \\
050924.05--672124.1 & SC25    & 0.26 & 20.27 & (2) \\
051853.19--690217.7 & SC7\phn & 1.63 & 20.13 &     \\
\enddata
\tablenotetext{a}{OGLE ID contains J2000.0 equatorial coordinates.}
\tablenotetext{b}{(1) Variable in \.Zebru\'n et al.\ 2001; (2) AGN
host galaxy visible.}
\end{deluxetable}


\begin{deluxetable}{ccccccccc}
\tabletypesize{\scriptsize}
\tablewidth{0pt}
\tablecaption{Quasar X-ray data.\label{tab:x}}
\tablehead{%
\colhead{Coords.\tablenotemark{a}} &
\colhead{ObsID\tablenotemark{b}} &
\colhead{Exp\tablenotemark{c}} &
\colhead{$N_X$\tablenotemark{d}} &
\colhead{$N_{\rm H,ATCA}$\tablenotemark{e}} &
\colhead{$N_{\rm H,X}$\tablenotemark{f}} &
\colhead{$\Gamma$\tablenotemark{g}} &
\colhead{Norm\tablenotemark{h}} &
\colhead{Remarks\tablenotemark{i}}
}
\startdata
05 07 36.30 $-$68 47 51.9 & 125 & 37.2 & 340 & 1.99 &
 $<6.0$        & 1.57$\pm$0.69 & 1.7$\pm$1.0 & (1) \\
05 08 33.18 $-$68 54 27.9 & 125 & 37.2 & 890 & 2.42 &
 1.47$\pm$0.57 & 2.19$\pm$0.24 & 7.3$\pm$1.4 & (2) \\
05 09 23.94 $-$67 21 23.6 & 776 & 49.6 & 440 & 2.27 &
 1.93$\pm$0.93 & 2.24$\pm$0.41 & 2.8$\pm$0.9 & (3) \\
05 18 53.02 $-$69 02 17.6 & 118 & 39.7 & 120 & 1.81 &
 $<8.3$        & 2.08$\pm$0.76 & 0.6$\pm$0.5 & \\
\enddata
\tablenotetext{a}{Chandra J2000.0 equatorial coordinates. Uncertainty
is $\sim$1~arcsec.}
\tablenotetext{b}{Chandra observation ID.}
\tablenotetext{c}{Chandra exposure time, in ks.}
\tablenotetext{d}{Net source events after background subtraction.}
\tablenotetext{e}{LMC HI column density, in $10^{21}$ cm$^{-2}$, from
ATCA observations by Kim et al 1998. Values are $\pm0.01$.}
\tablenotetext{f}{LMC absorbing column, in $10^{21}$ cm$^{-2}$, from
spectral fit. ``$<$'' indicates 3-$\sigma$ upper limit.}
\tablenotetext{g}{Photon spectral index, from spectral fit.}
\tablenotetext{h}{Power law normalization at 1keV,
in $10^{-5}$ photons cm$^{-2}$s$^{-1}$keV$^{-1}$, from spectral fit.}
\tablenotetext{i}{Haberl \& Pietsch 1999 and Sasaki et al.\ 2000 info:
(1) [HP99] 724, [SHP2000] 35, classified as ``hard'' (but not as ``AGN'');
(2) [HP99] 756, [SHP2000] 40; (3) [HP99] 523.}
\end{deluxetable}


Out of 35 CXO/OGLE objects observed in January 2002, four turned out
to be new quasars. In principle, this gives a 11\%\ efficiency for the
method for searching for quasars, but this value is, of course,
subject to low number statistics. Table~\ref{tab:opt} contains a
summary of the optical properties of the four quasars. We note that
during the observation we could clearly identify the host galaxy for
OGLE050924.05--672124.1, the one with lowest emission redshift
($\zem=0.26$).

We present the optical spectra of our quasars in
Figure~\ref{fig:magellan}. In all of them, several emission lines are
clearly visible, allowing unambiguous determinations of redshifts.
All spectra show typical blue continua. One somewhat unusual feature
is the apparent lack of [O~{\sc iii}] emission lines in
OGLE050736.52--684751.7 (upper left panel in Fig.~\ref{fig:magellan}),
but at $\zem=0.53$ they may have been affected by the strong O$_2$
atmospheric band, which is clearly visible in the spectra.

We note that only one of the quasars, OGLE050833.29--685427.5, has
been identified by OGLE as a variable. Its light curve clearly shows
variability with the amplitude of $\sim$0.4~mag. Post factum, we
examined the light curves (provided to us by A.~Udalski) of the other
objects. The quasars are relatively faint, and the light curves are
rather noisy, although they qualitatively suggest that the objects may
be variable.

Table~\ref{tab:x} contains a summary of the X-ray properties of the
objects. We analyzed the X-ray data using tools from CIAO~2.2 and
SHERPA. We reprocessed the observations using Chandra calibration
database (CALDB) ver.~2.11. That allowed us to get corrected X-ray
positions of sources, and we found them to agree well with the optical
positions. We note that the positions of three of the quasars coincide
with unidentified X-ray sources in the lists of ROSAT LMC X-ray
sources from Haberl \& Pietsch (1999) and Sasaki, Haberl, \& Pietsch
(2000); see Table~\ref{tab:x}.

The net number of X-ray events collected from four quasars range
between $\sim$100 and $\sim$900, which is sufficient to establish
basic X-ray properties of the sources. In addition to that, one can
take advantage of the location of the quasars and --- at least for the
strongest of our quasars --- attempt to estimate the absorbing column
in the LMC.

We extracted source and background spectra from regions around the
quasars using {\em dmextract} and then performed spectral fits with
SHERPA. Pile-up in ACIS is not a problem here, both because the source
count rates are low and because all sources are far off-axis. We
excluded data with $E<0.4$~keV, since there are large calibration
uncertainties for soft X-rays.

We fitted each spectrum assuming the intrinsic quasar spectrum to be a
power law. We assumed fixed Galactic absorption towards the LMC,
$N_H=5.5\times10^{21}$~cm$^{-2}$ (Schwering \& Israel 1991), and we
allowed for additional absorption from the LMC. In all cases, this
model gave adequate fits. The derived spectral X-ray properties
(Tab.~\ref{tab:x}) are within typical ranges for quasars (e.g.\ Fiore
et al.\ 1998; Reeves \& Turner 2000).

We got a meaningful measurement of LMC absorption only for the two
strongest X-ray sources among our four quasars. For the two weaker
sources, the uncertainty in the LMC absorbing column was larger than
the derived value; for those objects we list the $3\sigma$ upper
limits in Table~\ref{tab:x}. For comparison, in Table~\ref{tab:x} we
list the hydrogen column density from the Australia Telescope Compact
Array (ATCA) 21~cm observations by Kim et al.\ (1998). The values
derived from the X-ray spectral fits are in reasonable agreement with
the measurements of the hydrogen content in the LMC. Our value for
OGLE050833.29--685427.5 differs from the ATCA result by $\sim2\sigma$,
but there are several factors that can readily explain the
difference. First, the assumed value for the Galactic absorption may
be overestimated, resulting in underestimating the LMC component of
the absorbing column. Second, it is well established (e.g.\ Kim et
al.\ 1998) that the distribution of hydrogen in the LMC is far from
homogeneous. The measurements from Kim et al.\ (1998) are effectively
averaged over spatial resolution element of $\sim$1~arcmin (15~pc at
the LMC), while the X-ray absorption is probing a specific line of
sight.

\section{Summary and discussion}

We identified four previously unknown quasars behind the Large
Magellanic Cloud, including one quasar at relatively high redshift,
$\zem=1.63$, by combining publicly available Chandra and OGLE data and
ground-based followup spectroscopy. While the underlying idea in this
project was to search for quasars based on {\em both} their X-ray
positions {\em and} variability, we note that only one of our four
quasars was positively identified by OGLE as a variable object. The
validity of our assumption that the combined Chandra/OGLE pointings
are good enough to accurately pinpoint quasar positions is an
excellent starting point for future studies of X-ray selected quasars
behind the LMC. So far, we have found four quasars in
$\sim$0.5~deg$^2$ covered by the four analyzed Chandra observations,
which compares favorably with $\sim$30 quasars in $\sim$11~deg$^2$,
discovered by the MACHO project on the basis of their variability

The intrinsic X-ray and optical properties of all four quasars are
quite typical. All four quasars are relatively faint ($V=19-20$),
which, at least at present, makes them unlikely targets for
high-resolution observations, especially with the Hubble Space
Telescope. HST observations would be necessary for the analysis of the
most important absorption lines, such as Ly-$\alpha$, C~{\sc iv} or
Mg~{\sc ii}, which for LMC occur in the UV part of the spectrum, below
the atmospheric break. However, the quasars will be very attractive
targets for the new generation of large telescopes, as well as the
NGST.

Three of the quasars are well positioned to become reference points
for proper motion studies. They are in dense fields:
OGLE050736.52--684751.7 and OGLE050833.29--685427.5 are near the edge
of the LMC bar and OGLE051853.19--690217.7 is near the center of the
bar. The fourth quasar, OGLE050924.05--672124.1, is outside of the LMC
bar, and it is also the one for which the host galaxy is visible,
limiting somewhat its application for this type of project.

We add that the followup spectroscopy of the CXO/OGLE candidates
revealed several other interesting objects, such as very hot stars,
X-ray binaries, etc. We will present the analysis of those objects in
a forthcoming paper.


\acknowledgments

We would like to thank B.~Paczy\'nski, A.~Siemiginowska and the
referee for helpful comments. We would also like to thank K.~Borkowski
for allowing us to use his Chandra observation prior to it becoming
public, S.~Kim and A.~Udalski for providing us with unpublished data,
and T.~Anselowitz for help with processing the X-ray observations. We
are grateful to the OGLE collaboration for putting their large data
sets in public domain. This research has made use of the NASA/IPAC
Extragalactic Database (NED) which is operated by the Jet Propulsion
Laboratory, California Institute of Technology, under contract with
NASA, and of the SIMBAD database, operated at CDS, Strasbourg,
France. AD acknowledges support from NASA Contract No.\ NAS8-39073
(CXC). PJG was supported by a CfA fellowship.


\end{document}